\documentclass[12pt]{article} 

\advance\textwidth by +1.0in 
\advance\textheight by +1.0in 
\advance\oddsidemargin by -0.5in 
\advance\evensidemargin by -1.0in 
\advance\topmargin by -0.5in 
\def\ov#1{\overline{#1}} 
 
\def\wt#1{\widetilde{#1}} 
\def\vb#1{\mbox{\boldmath$#1$}} 
\def\pd#1#2{\frac{\partial #1}{\partial #2}} 
 
\def\wh#1{\widehat{#1}} 
\def\bdot{\,\vb{\cdot}\,} 
\def\btimes{\,\vb{\times}\,}

\newcommand{\bc}{\begin{center}} 
\newcommand{\ec}{\end{center}} 
\newcommand{\bt}{\begin{tabbing}} 
\newcommand{\et}{\end{tabbing}} 
\newcommand{\be}{\begin{eqnarray*}} 
\newcommand{\ee}{\end{eqnarray*}} 
 
\newcommand{\no}{\noindent} 

\begin{document} 

\bc 
{\Large {\sf Relativistic quasilinear diffusion in axisymmetric magnetic geometry for arbitrary-frequency electromagnetic fluctuations}} 

\vspace*{0.4in} 

Alain J. Brizard \\ 
{\it Department of Chemistry and Physics, Saint Michael's College} \\ 
{\it Colchester, Vermont 05439} \\ 

\vspace*{0.1in} 

and 

\vspace*{0.1in} 

Anthony A. Chan \\ 
{\it Department of Physics and Astronomy, Rice University} \\ 
{\it Houston, Texas 77005} 
\ec 

\vspace*{0.5in} 

A relativistic bounce-averaged quasilinear diffusion equation is derived to describe stochastic particle transport associated with arbitrary-frequency electromagnetic fluctuations in a nonuniform magnetized plasma. Expressions for the elements of a relativistic quasilinear diffusion tensor are calculated explicitly for magnetically-trapped particle distributions in axisymmetric magnetic geometry in terms of gyro-drift-bounce wave-particle resonances. The resonances can destroy any 
one of the three invariants of the unperturbed guiding-center Hamiltonian dynamics. 

\vfill  

\no 

\vfill\eject 

\no 
{\sf I. INTRODUCTION} 

\vspace*{0.2in} 

Understanding the phase-space transport of magnetically-trapped relativistic electrons is an intrinsically interesting general problem, and it is an especially important problem in magnetospheric plasma physics because these energetic particles can damage spacecraft electronics and they present a radiation hazard to astronauts. The quasilinear theory of such anomalous transport processes is based on resonant wave-particle interactions in which characteristic wave frequencies match one or more of the orbital frequencies associated with the gyration, bounce, and drift motion of charged particles \cite{Kaufman,schulz74,mynick89}. In particular, diffusion due to drift-resonances with low-frequency magnetohydrodynamic (MHD) waves \cite{hudson98,elkington99,elkington03} and diffusion due to cyclotron-resonances with various high-frequency waves \cite{lyons71,lyons74,albert99,horne03a} are frequently cited as important transport mechanisms for relativistic electrons in Earth's magnetosphere. Previous work on these transport mechanisms has typically been restricted to either the low-frequency wave interactions, which may break the second and third invariants, or the high-frequency wave interactions, which may break the first and second invariants. However, there has been little work on a unified theory in which all the interactions can be considered together. 

Our previous work \cite{BC_2001} (henceforth refered to as Paper I) investigated relativistic quasilinear diffusion transport due to low-frequency electromagnetic fluctuations, which preserved the first adiabatic invariant, based on the low-frequency relativistic gyrokinetic Vlasov equation \cite{tsai84,chen99,BC_1999}. In the present work, we modify the low-frequency gyrokinetic formalism to allow for arbitrary-frequency electromagnetic fluctuations (and, thus, allow for processes which break any one of the three invariants) and retain full finite-Larmor-radius (FLR) effects associated with the electromagnetic fluctuations. Although FLR effects are not important for interactions of relativistic electrons with MHD waves, they are crucial for interactions with cyclotron-frequency waves 
\cite{lyons71,lyons74}. 

\vspace*{0.2in} 

\no 
{\bf A. Canonical Quasilinear Formalism} 

\vspace*{0.2in} 

The derivation of a quasilinear diffusion equation in canonical action space (with coordinates ${\bf J}$) was first performed by Kaufman \cite{Kaufman}, in which the background Vlasov distribution $F_{0}({\bf J},\tau)$ evolves on a slow time scale $(\tau = \epsilon^{2}\,t)$ determined by the amplitude (ordered by the dimensionless parameter $\epsilon$) of the electric and magnetic fluctuations. The quasilinear diffusion equation in action space is given in its general form as 
\begin{equation} 
\pd{F_{0}}{\tau} \;\equiv\; \pd{}{{\bf J}}\bdot \left( {\sf D}_{{\rm CQL}} \bdot\pd{F_{0}}{{\bf J}} \right), 
\label{eq:QL_eq} 
\end{equation} 
where the action coordinates ${\bf J}$ may either be {\it exact} invariants (corresponding to exact symmetries) or {\it adiabatic} invariants (corresponding to approximate symmetries). The canonical quasilinear (CQL) diffusion tensor \cite{Kaufman,BC_2001} is expressed as 
\begin{equation} 
{\sf D}_{{\rm CQL}} \;=\; \sum_{{\bf m},k}\; {\bf m}{\bf m}\; \left[\;\pi\;\delta(\omega_{k} - {\bf m}\bdot \vb{\Omega})\;|\delta\wt{H}_{{\bf m}k}|^{2}\;\right], 
\label{eq:QL_Diff} 
\end{equation} 
where $\delta\wt{H}_{{\bf m}k}$ denotes the Fourier component of the perturbation Hamiltonian $\delta H$ with a discrete frequency spectrum (represented by the wave frequency $\omega_{k}$) and ${\bf m}\bdot\vb{\Omega} = m^{i}\, \Omega_{i}$ (summation over repeated indices is, henceforth, implied), with $\Omega_{i} = \partial H_{0}/\partial J^{i}$ denoting the unperturbed orbital frequency derived from a Hamiltonian $H_{0}$ averaged over all fast orbital time scales. 

The expression (\ref{eq:QL_Diff}) clearly establishes the paradigm of quasilinear transport theory: wave-particle resonances (where $\omega_{k} = {\bf m}\bdot
\vb{\Omega}$) introduce explicit violations of the invariance (exact or adiabatic) of action coordinates leading to stochastic transport in action space. Note that the canonical quasilinear diffusion tensor (\ref{eq:QL_Diff}) can be written 
\begin{equation} 
{\sf D}_{{\rm CQL}} \;=\; \sum_{{\bf m},k}\; {\sf d}_{{\bf m},k}^{({\rm CQL})}\;\Gamma_{{\bf m},k}^{({\rm CQL})}, 
\label{eq:QL_dGamma} 
\end{equation} 
in terms of the canonical quasilinear {\it coefficient matrix} ${\sf d}_{{\bf m},k}^{({\rm CQL})} = {\bf m}{\bf m}$ and the canonical quasilinear {\it potential} 
$\Gamma_{{\bf m},k}^{(CQL)} = \pi\;\delta(\omega_{k} - {\bf m}\bdot\vb{\Omega})\;|\delta\wt{H}_{{\bf m}k}|^{2}$. Here, the {\it dimensionality} of quasilinear transport is represented by the dimensionality of the coefficient matrix ${\sf d}_{{\bf m},k}$ and the {\it universality} of quasilinear transport is represented by the fact 
that a single quasilinear potential $\Gamma_{{\bf m},k}$ describes quasilinear transport along different action coordinates. 

\vspace*{0.2in} 

\no 
{\bf B. Organization of Paper} 

\vspace*{0.2in} 

The remainder of the paper is organized as follows. In Sec.~II, we introduce the unperturbed and perturbed relativistic guiding-center equations in terms 
of magnetic coordinates $(\psi,\varphi,s)$ associated with an unperturbed axisymmetric magnetic field configuration. The invariant coordinates ${\bf I} = (J_{{\rm g}}, \varepsilon, J_{{\rm d}})$ in axisymmetric magnetic geometry are the relativistic guiding-center gyro-action $J_{{\rm g}}$, the relativistic guiding-center kinetic energy $\varepsilon$, and the guiding-center drift action $J_{{\rm d}} = q\psi/c$. Here, energy $\varepsilon$ is used instead of the bounce action $J_{{\rm b}}$ to ensure that all three of the invariant coordinates are {\it local} (i.e., they depend on local properties of the background plasma). The perturbation 
guiding-center Hamiltonian $\delta H$, expressed in terms of the perturbation electromagnetic potentials $(\delta\phi,\,\delta{\bf A})$, causes the 
destruction of the unperturbed invariants ${\bf I}$ (i.e., $\delta\dot{{\bf I}} = \{ {\bf I},\; \delta H\} \neq 0$, where $\{\,,\,\}$ represents the 
Poisson bracket) and leads to stochastic quasilinear diffusion in invariant ${\bf I}$-space due to resonant wave-particle resonances. 

In Secs.~III and IV, an explicit expression for the relativistic quasilinear diffusion equation (\ref{eq:QL_eq}) in axisymmetric magnetic geometry in terms of the invariant coordinates $(J_{{\rm g}}, \varepsilon, J_{{\rm d}})$ is derived.  In Sec.~III, the quasilinear coefficient matrix ${\sf d}_{{\bf m}k}$ is derived and then, in Sec.~IV, the quasilinear potential $\Gamma_{{\bf m},k}$. The present derivation follows closely the derivation found in Paper I \cite{BC_2001}. We summarize our results in Sec.~V and discuss applications. 

Since quasilinear diffusion is often described in the literature (e.g., see Ref.~\cite{schulz74}) in terms of the equatorial pitch angle instead of the gyro-action, the corresponding relativistic quasilinear diffusion equation (\ref{eq:QL_eq}) is presented in Appendix A. In particular, we show how the quasilinear coefficient matrix transforms when the relativistic guiding-center gyro-action $J_{{\rm g}}$ is replaced with the equatorial pitch angle.  Lastly, Appendix B presents two expressions for 
the bounce-averaged drift frequency which might be useful for computational applications. 

\vspace*{0.2in} 

\bt 
{\sf II.} \={\sf RELATIVISTIC GUIDING-CENTER DYNAMICS IN AXISYMMETRIC MAGNETIC} \\ 
          \>{\sf GEOMETRY} 
\et 

We begin our analysis of relativistic quasilinear transport in axisymmetric magnetic geometry by first providing a general representation of axisymmetric magnetic geometry in terms of magnetic coordinates (also known as Euler or Clebsch potentials). Next, we present the equations of relativistic guiding-center Hamiltonian dynamics in unperturbed and perturbed axisymmetric magnetic geometry.  Unperturbed guiding-center Hamiltonian dynamics establishes the existence of three invariants, which are then destroyed by arbitrary-frequency electromagnetic perturbation fields. 

In the spirit of perturbation theory, we assume that electric fields are not part of the quasi-static background fields (i.e., electric fields are automatically viewed as perturbation fields) and any axisymmetry-breaking magnetic fields are viewed as zero-frequency magnetic perturbation fields. The validity of these assumptions will be checked in future work when an alternate model including a more realistic non-axisymmetric magnetic geometry and a background electric field is considered. 

\vspace*{0.2in} 

\no 
{\bf A.  Axisymmetric Magnetic Geometry} 

\vspace*{0.2in} 

The background magnetic field in axisymmetric magnetic geometry can be expressed in terms of magnetic coordinates $(\psi,\varphi,s)$ as 
\begin{equation} 
{\bf B} \;=\; \nabla\psi\btimes\nabla\varphi \;=\; B(\psi,s)\;\pd{{\bf X}}{s}, 
\label{eq:axi_mag} 
\end{equation} 
where $\psi$ denotes the magnetic (radial) flux function (i.e., ${\bf B} \bdot\nabla\psi = 0)$, the azimuthal angle $\varphi$ is an ignorable coordinate (i.e., in axisymmetric magnetic geometry, unperturbed scalar fields are independent of $\varphi$), and $s$ is the parallel spatial coordinate along a single magnetic field line labeled by $(\psi,\varphi)$. 

The magnitude $B(\psi,s)$ of the magnetic field, on the one hand, is defined from Eq.~(\ref{eq:axi_mag}) as $B = \nabla\psi\btimes \nabla\varphi\bdot\nabla s$ and 
$B^{-1}(\psi,s)$ denotes the Jacobian for the transformation ${\bf X} \rightarrow (\psi,\varphi,s)$. The local magnetic unit vector $\wh{{\sf b}} = {\bf B}/B = 
\partial_{s}{\bf X}$, on the other hand, can be expressed as 
\begin{equation} 
\wh{{\sf b}} \;=\; \nabla s \;+\; a(\psi,s)\;\nabla\psi \, 
\label{eq:b_hat} 
\end{equation} 
Here the scalar field 
\[ a(\psi,s) \;=\; \wh{{\sf b}}\bdot\pd{{\bf X}}{\psi} \;=\; -\;\frac{\nabla\psi\bdot\nabla s}{|\nabla\psi|^{2}}, \] 
which characterizes the non-orthogonality of the coordinates $(\psi, s)$, is required to ensure that $\nabla\btimes\wh{{\sf b}} \neq 0$ and is associated with 
magnetic curvature  \cite{BC_2001} through the relation $\partial_{s}\wh{{\sf b}} = (\partial_{s}a)\,\nabla\psi$. We note that, for an axisymmetric dipole magnetic 
field (which is curl-free), we can write ${\bf B} = \nabla\chi(\psi,s)$ and, thus, using Eq.~(\ref{eq:b_hat}), we find $\partial\chi/\partial s = B$ and 
$\partial\chi/\partial\psi = aB$. 

Next, we define the radius $R(\psi,s) \equiv |\nabla\varphi|^{-1}$ (which measures the distance to the symmetry axis) and obtain the expression $|\nabla\psi| = BR$ from Eq.~(\ref{eq:axi_mag}). We can, thus, define the following perpendicular unit vectors 
\begin{equation} 
\wh{\psi} \;=\; (BR)^{-1}\;\nabla\psi \;\;\;{\rm and}\;\;\; \wh{\varphi} \;=\; R\;\nabla\varphi, 
\label{eq:unit} 
\end{equation} 
such that the unit vectors $(\wh{\psi},\wh{\varphi},\wh{{\sf b}})$ form a right-handed set of unit vectors (i.e., $\wh{{\sf b}} = \wh{\psi}\btimes\wh{\varphi}$). 

\vspace*{0.2in} 

\no 
{\bf B.  Unperturbed Guiding-Center Dynamics} 

\vspace*{0.2in} 

The unperturbed relativistic guiding-center dynamics of a charged particle of mass $M$ and charge $q$ in axisymmetric magnetic geometry is represented in terms of the magnetic coordinates $(\psi,\varphi,s)$, the unperturbed relativistic guiding-center kinetic energy 
\begin{equation} 
\varepsilon \;=\; \left( \gamma - 1 \left)\,Mc^{2} \right.\right. \;=\; \sqrt{M^{2}c^{4} \;+\; 2\,Mc^{2}\,J_{{\rm g}} \omega_{{\rm g}} \;+\; p_{\|}^{2}c^{2}} \;-\; 
Mc^{2}, 
\label{eq:energy_gc} 
\end{equation} 
where $\omega_{{\rm g}} = qB/Mc$ denotes the rest-mass gyro-frequency and $p_{\|} = \gamma\,Mv_{\|}$ denotes the parallel component of the relativistic kinetic momentum, the relativistic guiding-center gyro-action $J_{{\rm g}} = |{\bf p}_{\bot} |^{2}/(2\,M\omega_{{\rm g}})$, and the gyro-angle $\zeta$. The unperturbed relativistic guiding-center Lagrangian is written as 
\begin{equation} 
L_{0} \;=\; J_{{\rm d}}\;\dot{\varphi} \;+\; p_{\|}\;\dot{s} \;+\; J_{{\rm g}}\;\dot{\zeta} \;-\; \varepsilon, 
\label{eq:Lag_unp} 
\end{equation} 
where ${\bf A} = \psi\,\nabla\varphi$ is the vector potential associated with the magnetic field (\ref{eq:axi_mag}) and $J_{{\rm d}} = q\psi/c$ denotes the drift action, while, using Eq.~(\ref{eq:energy_gc}), the magnitude of the relativistic guiding-center parallel momentum is 
\begin{equation} 
|p_{\|}| \;=\; \gamma M\,|v_{\|}| \;=\; \sqrt{\varepsilon^{2}/c^{2} \;+\; 2M\left( \varepsilon \;-\; J_{{\rm g}} \omega_{{\rm g}}\right)}. 
\label{eq:p_par} 
\end{equation}  
The unperturbed relativistic guiding-center Poisson bracket $\{\;,\;\}$, on the other hand, is expressed in terms of two arbitrary functions $F$ and $G$ on guiding-center phase space as 
\begin{eqnarray} 
\{ F,\; G \} & = & \pd{F}{\zeta} \left( \pd{G}{J_{{\rm g}}} \;+\; \omega_{{\rm c}}\;\pd{G}{\varepsilon} \right) \;-\; \left( \pd{F}{J_{{\rm g}}} \;+\; 
\omega_{{\rm c}}\;\pd{F}{\varepsilon} \right) \pd{G}{\zeta} \nonumber \\ 
 &   &\mbox{}+\; \pd{F}{\varphi} \left( \pd{G}{J_{{\rm d}}} \;+\; \omega_{{\rm d}}\;\pd{G}{\varepsilon} \;-\; \frac{ca}{q}\;\pd{G}{s} \right) \;-\; 
\left( \pd{F}{J_{{\rm d}}} \;+\; \omega_{{\rm d}}\;\pd{F}{\varepsilon} \;-\; \frac{ca}{q}\;\pd{F}{s} \right) \pd{G}{\varphi} \nonumber \\ 
 &   &\mbox{}+\; v_{\|} \left( \pd{F}{s}\;\pd{G}{\varepsilon} \;-\; \pd{F}{\varepsilon}\;\pd{G}{s} \right), 
\label{eq:Poisson_unp} 
\end{eqnarray} 
where $\omega_{{\rm c}} \equiv \gamma^{-1}\,\omega_{{\rm g}}$ denotes the relativistic gyro-frequency and the azimuthal drift frequency is defined as 
\begin{equation} 
\omega_{{\rm d}} \;=\; J_{{\rm g}} \left( \pd{\omega_{{\rm c}}}{J_{{\rm d}}} \;-\; \frac{ac}{q}\; \pd{\omega_{{\rm c}}}{s} \right) \;+\; 
\frac{p_{\|}^{2}}{\gamma\,M}\;\pd{a}{s}. 
\label{eq:axi_drift} 
\end{equation} 
The Jacobian for the guiding-center transformation 
\[ ({\bf x},{\bf p}) \;\rightarrow\; Z^{\alpha} \;=\; (J_{{\rm d}},\varphi,s,\varepsilon,J_{{\rm g}},\zeta) \] 
is ${\cal J} = 1/|v_{\|}|$, where $|v_{\|}|$ is obtained from Eq.~(\ref{eq:p_par}). We note that the Poisson bracket (\ref{eq:Poisson_unp}) can also be expressed in phase-space divergence form as 
\begin{equation} 
\{ F,\; G\} \;=\; \pd{F}{Z^{\alpha}}\;J^{\alpha\beta}\;\pd{G}{Z^{\beta}} \;=\; \frac{1}{{\cal J}}\; \pd{}{Z^{\alpha}} \left( {\cal J}\,F\;J^{\alpha\beta}\; 
\pd{G}{Z^{\beta}} \right), 
\label{eq:PB_identity} 
\end{equation} 
where $J^{\alpha\beta} = \{ Z^{\alpha},\; Z^{\beta}\}$ denotes the elements of the antisymmetric Poisson-bracket tensor and we used the Liouville identities ($\beta = 1,...,6)$
\begin{equation} 
\pd{}{Z^{\alpha}}\left( {\cal J}\;J^{\alpha\beta} \right) \;=\; 0 . 
\label{eq:Liouville} 
\end{equation} 
These identities follow from the divergenceless property of the Hamiltonian flow (associated with Liouville's Theorem): 
\begin{eqnarray*} 
0 & = & \pd{}{{\bf Z}}\bdot\dot{{\bf Z}} \;=\; \frac{1}{{\cal J}}\;\pd{}{Z^{\alpha}} \left( {\cal J}\;\dot{Z}^{\alpha} \right) \;=\; \frac{1}{{\cal J}}\;
\pd{}{Z^{\alpha}} \left( {\cal J}\;J^{\alpha\beta}\;\pd{H}{Z^{\beta}} \right) \\ 
 & = & \frac{1}{{\cal J}}\;\pd{}{Z^{\alpha}} \left( {\cal J}\;J^{\alpha\beta} \right)\;\pd{H}{Z^{\beta}} \;+\; J^{\alpha\beta}\;\frac{\partial^{2}H}{\partial 
Z^{\alpha}\partial Z^{\beta}}. 
\end{eqnarray*} 
Since the Poisson matrix $J^{\alpha\beta}$ is anti-symmetric, the term $J^{\alpha\beta}\,\partial_{\alpha\beta}^{2}H$ vanishes, and we find 
\[ 0 \;=\; \frac{1}{{\cal J}}\;\pd{}{Z^{\alpha}} \left( {\cal J}\;J^{\alpha\beta} \right)\;\pd{H}{Z^{\beta}}, \] 
which must be true for all Hamiltonians $H$ and, thus, we obtain the Liouville identities (\ref{eq:Liouville}). In particular, the Liouville identities for $\varphi$ and $\zeta$, respectively, are  
\begin{eqnarray} 
\pd{}{J_{{\rm d}}}\left(\frac{1}{|v_{\|}|}\right) \;-\;  \pd{}{s}\left(\frac{ca}{q|v_{\|}|}\right) \;+\; \pd{}{\varepsilon}\left(\frac{\omega_{{\rm d}}}{|v_{\|}|}
\right) & = & 0, \nonumber \\ 
 &  & \label{eq:ids} \\ 
\pd{}{J_{{\rm g}}}\left(\frac{1}{|v_{\|}|}\right) \;+\; \pd{}{\varepsilon}\left(\frac{\omega_{{\rm c}}}{|v_{\|}|}\right)  & = & 0, \nonumber 
\end{eqnarray} 
which follow from the definitions for $\omega_{{\rm c}}$, $|v_{\|}|$, and $\omega_{{\rm d}}$. We further note that although the magnetic coordinate $\psi$ and the drift action $J_{{\rm d}}$ are simply related in axisymmetric magnetic geometry, this simple relation is destroyed by non-axisymmetry; we, henceforth, use $\psi$ and 
$J_{{\rm d}}$ separately as follows: $J_{{\rm d}}$ is used whenever dynamics is concerned while $\psi$ is used whenever magnetic coordinates are concerned.  

As in Paper I \cite{BC_2001}, the unperturbed guiding-center Hamilton's equations, expressed in terms of the coordinates $(J_{{\rm d}},\varphi,s;\; \varepsilon, 
J_{{\rm g}},\zeta)$ and the unperturbed relativistic guiding-center Hamiltonian $H_{0} = \varepsilon$, include the magnetic-coordinate components of the relativistic guiding-center velocity 
\begin{equation} 
\dot{J}_{{\rm d}0} \;=\; 0,\;\; \dot{\varphi}_{0} \;=\; \omega_{{\rm d}}, \;\;{\rm and}\;\; \dot{s}_{0} \;=\; v_{\|} 
\label{eq:axi_x} 
\end{equation} 
and the velocity-space guiding-center equations 
\begin{equation} 
\dot{\varepsilon}_{0} \;=\; 0 \;=\; \dot{J}_{{\rm g}0} \;\;\;{\rm and}\;\;\; \dot{\zeta}_{0} \;=\; \omega_{{\rm c}} \, . 
\label{eq:ugc_zeta} 
\end{equation} 
Thus, the drift action $J_{{\rm d}} \;(= q\psi/c)$, the kinetic energy $\varepsilon$, and the gyro-action $J_{{\rm g}}$ are the three invariants of unperturbed relativistic guiding-center motion in axisymmetric magnetic geometry. From these unperturbed Hamilton's equations we construct the unperturbed time evolution (Vlasov) operator 
\begin{equation} 
\frac{d_{0}}{dt} \;\equiv\; \pd{}{t} \;+\; v_{\|}\;\pd{}{s} \;+\; \omega_{{\rm d}}\;\pd{}{\varphi} \;+\; \omega_{{\rm c}}\;\pd{}{\zeta}. 
\label{eq:d0_dt} 
\end{equation} 
In the absence of electromagnetic fluctuations, the unperturbed Vlasov equation $d_{0}F_{0}/dt = 0$, therefore, implies that the unperturbed (background) 
Vlasov distribution $F_{0}({\bf I})$ is a function of the three guiding-center invariants ${\bf I} = (J_{{\rm g}},\varepsilon,J_{{\rm d}})$ associated with axisymmetric magnetic geometry. In the present work, in order to correctly account for wave-particle gyroresonance effects, we retain full finite-Larmor-radius (FLR) effects 
associated with electromagnetic fluctuations and, thus, we need an expression for the gyroradius $\vb{\rho}$. The gyroradius vector $\vb{\rho}({\bf X},\varepsilon, 
J_{{\rm g}},\zeta)$ is defined locally in terms of the magnetic coordinates as 
\begin{equation} 
\vb{\rho} \;=\; \rho \left( \cos\zeta\;\wh{\psi} \;-\; \sin\zeta\;\wh{\varphi} \right), 
\label{eq:gyroradius} 
\end{equation} 
where the unit vectors $(\wh{\psi},\,\wh{\varphi})$ are defined in Eq.~(\ref{eq:unit}) and the magnitude $\rho = v_{\bot}/\omega_{{\rm c}}$ of the gyro-radius vector 
$\vb{\rho}$ is given as 
\begin{equation} 
\rho(J_{{\rm g}};\;\psi,s) \;=\; \sqrt{\frac{2\,J_{{\rm g}}}{M\omega_{{\rm g}}(\psi,s)}}. 
\end{equation} 

In Sec.IV, the spatial dependence of the perturbed electromagnetic fields (in guiding-center phase- space) will involve the combination $\psi^{i} + \vb{\rho}\bdot\nabla\psi^{i}$, which denotes the position ${\bf x} = {\bf X} + \vb{\rho}$ of a particle in terms of the guiding-center magnetic coordinates $\psi^{i} = (\psi,\varphi,s)$. Here, using Eqs.~(\ref{eq:unit}) and (\ref{eq:gyroradius}), we find 
\begin{equation} 
\left. \begin{array}{rcl} \vb{\rho}\bdot\nabla\psi & = & \rho\,|\nabla\psi|\;\cos\zeta \;\equiv\; \Delta\psi\;\cos\zeta \\
 & & \\ 
\vb{\rho}\bdot\nabla\varphi & = & -\;\rho\,|\nabla\varphi|\;\sin\zeta \;\equiv\; -\;(\rho/R)\;\sin\zeta \\
 & & \\ 
\vb{\rho}\bdot\nabla s & = & -\;a\;\vb{\rho}\bdot\nabla\psi \;\equiv\; -\;a\;\Delta\psi\;\cos\zeta, \end{array} \right\}, 
\label{eq:rho_grad} 
\end{equation} 
so that a generic perturbation scalar field 
\begin{equation} 
\delta\chi({\bf X} + \vb{\rho}, t) \;=\; \exp(\vb{\rho}\bdot\nabla)\,\delta\chi({\bf X}, t) \;\equiv\; \delta_{gc}\chi( \psi,\varphi,s;\; \varepsilon,J_{{\rm g}},
\zeta, t) 
\label{eq:chi_gc} 
\end{equation} 
can be expressed as 
\begin{equation} 
\delta\chi({\bf X} + \vb{\rho}, t) \;=\; \exp\left[\; -\;\left(\frac{\rho}{R}\,\sin\zeta\right)\;\pd{}{\varphi} \;+\; \Delta\psi\,\cos\zeta \left( \pd{}{\psi} \;-\; 
a\;\pd{}{s} \right)\;\right] \delta\chi(\psi,\varphi,s, t). 
\label{eq:Taylor_chi} 
\end{equation} 
This expression will be used later in obtaining Fourier-Bessel expansions in gyrorangle $\zeta$ for the electromagnetic perturbation fields. 

\vspace*{0.2in} 

\no 
{\bf C.  Perturbed Guiding-Center Dynamics} 

\vspace*{0.2in} 

The presence of fluctuating electric and magnetic fields 
\[ \delta{\bf E} \;=\; -\,\nabla\delta\phi \;-\; c^{-1}\,\partial_{t}\delta{\bf A} \;\;\;{\rm and}\;\;\; \delta{\bf B} \;=\; \nabla\btimes\delta{\bf A}, \] 
where $\delta\phi$ and $\delta{\bf A}$ are the perturbed scalar and vector potentials, respectively, implies that the three unperturbed invariants $(J_{{\rm g}}, 
\varepsilon,J_{{\rm d}})$ are no longer invariants. The perturbed electromagnetic potentials introduce perturbations in the relativistic guiding-center Lagrangian 
(\ref{eq:Lag_unp}): $L_{0} \rightarrow L_{0} + \delta L$, which, to first order in the perturbed potentials, yields 
\begin{equation} 
\delta L \;=\; \frac{q}{c}\,\delta_{{\rm gc}}{\bf A}\bdot \left( \dot{{\bf X}}_{0} \;+\; \dot{\vb{\rho}}_{0} \right) \;-\; q\,\delta_{{\rm gc}}\phi \,, 
\label{eq:Lag_p} 
\end{equation} 
where the notation $\delta_{gc}\chi$ is defined in Eq.~(\ref{eq:chi_gc}). As a result of the magnetic perturbation (the first term in Eq.~(\ref{eq:Lag_p})), 
the relativistic guiding-center Poisson bracket (\ref{eq:Poisson_unp}) is also perturbed unless we define the perturbed Hamiltonian $\delta H$ as 
\begin{equation} 
\delta H \;\equiv\; -\;\delta L \;=\; q \left( \delta_{{\rm gc}}\phi \;-\; \frac{v_{\|}}{c}\;\delta_{{\rm gc}}A_{\|} \right) \;-\; \frac{q}{c}\;{\bf v}_{\bot}
\bdot \delta_{{\rm gc}}{\bf A}_{\bot}. 
\label{eq:delta_Hgc} 
\end{equation} 
The destruction of the three unperturbed guiding-center invariants is expressed in terms of the perturbed guiding-center Hamiltonian $\delta H$ and the 
{\it unperturbed} Poisson bracket (\ref{eq:Poisson_unp}) as 
\begin{eqnarray} 
\delta\dot{J}_{{\rm g}} & = & \{ J_{{\rm g}},\; \delta H\} \;=\; -\;\pd{\delta H}{\zeta}, \label{eq:delta_g} \\ 
\delta\dot{\varepsilon} & = & \{ \varepsilon,\; \delta H \} \;=\; \left( \pd{}{t} - \frac{d_{0}}{dt} \right) \delta H, \label{eq:delta_e} \\ 
\delta\dot{J}_{{\rm d}} & = & \{ J_{{\rm d}},\; \delta H \} \;=\; -\;\pd{\delta H}{\varphi}, \label{eq:delta_d} 
\end{eqnarray} 
where the unperturbed time evolution operator $d_{0}/dt$ is defined by Eq.~(\ref{eq:d0_dt}), while the remaining perturbed guiding-center Hamilton's equations are 
\begin{eqnarray} 
\delta\dot{\zeta} & = & \{ \zeta,\; \delta H \} \;=\; \left( \pd{}{J_{{\rm g}}} \;+\; \omega_{{\rm c}}\;\pd{}{\varepsilon} \right) \delta H, 
\label{eq:delta_zeta} \\ 
\delta\dot{s} & = &  \{ s,\; \delta H \} \;=\; \left( v_{\|}\;\pd{}{\varepsilon} \;+\; \frac{ca}{q}\;\pd{}{\varphi} \right) \delta 
H, \label{eq:delta_s} \\ 
\delta\dot{\varphi} & = & \{ \varphi,\; \delta H \} \;=\; \left( \pd{}{J_{{\rm d}}} \;-\; \frac{ca}{q}\;\pd{}{s} \;+\; \omega_{{\rm d}}\; 
\pd{}{\varepsilon} \right) \delta H. \label{eq:delta_phi} 
\end{eqnarray} 
From Eqs.~(\ref{eq:delta_g}) and (\ref{eq:delta_d}), we note that the gyro-action $J_{{\rm g}}$ and the drift-action $J_{{\rm d}}$ are destroyed if the perturbation Hamiltonian $\delta H$ depends on the gyro-angle $\zeta$ and the azimuthal angle $\varphi$, respectively. Moreover, the destruction of the gyro-action invariance leads to the possible loss of trapped particles while the destruction of the drift-action invariance leads to radial transport. 

\vspace*{0.2in} 

\no 
{\sf III. RELATIVISTIC QUASILINEAR DIFFUSION TENSOR} 

\vspace*{0.2in} 

In this Section, we proceed with a two-time-scale analysis of the perturbed Vlasov equation 
\begin{equation} 
\epsilon^{2}\;\pd{F_{0}}{\tau} \;+\; \epsilon\;\frac{d_{0}\delta F}{dt} \;=\; -\; \epsilon\; \{( F_{0} \;+\; \epsilon\;\delta F),\; \delta H\}, 
\label{eq:Vlasov_start} 
\end{equation} 
where the unperturbed evolution operator $d_{0}/dt$ is defined in Eq.~(\ref{eq:d0_dt}) and the Vlasov distribution $F$ is decomposed as $F = F_{0}({\bf I}, \tau =\epsilon^{2}t) + \epsilon\,\delta F$. The fast-time-scale evolution equation shows, on the one hand, how the perturbed Vlasov distribution $\delta F$ evolves under the influence of electromagnetic perturbation fields. The slow-time-scale evolution equation, on the other hand, shows how the background distribution $F_{0}$ changes as 
a result of wave-particle resonances. Following our previous work (Paper I), we now proceed with the separation of the perturbed guiding-center Vlasov distribution 
\begin{equation} 
\delta F \;=\; \pd{F_{0}}{\varepsilon}\;\delta H \;+\; \delta G 
\label{eq:ad_nonad} 
\end{equation} 
in terms of the {\it adiabatic} part of $\delta F$ (the first term) and the {\it non-adiabatic} part of $\delta F$ (the second term), which explicitly represents resonant wave-particle effects. The fast-time evolution equation for the non-adiabatic part $\delta G$ is obtained from the linearized Vlasov equation 
$d_{0}\delta F/dt = -\,\delta \dot{{\bf I}}\bdot\partial F_{0}/\partial {\bf I}$ as 
\begin{eqnarray} 
\frac{d_{0}\delta G}{dt} & \equiv & \frac{d_{0}\delta F}{dt} \;-\; \pd{F_{0}}{\varepsilon}\;\frac{d_{0}\delta H}{dt}  \nonumber \\ 
 & = & \left( \pd{F_{0}}{J_{{\rm d}}}\;\pd{}{\varphi} \;+\; \pd{F_{0}}{J_{{\rm g}}}\;\pd{}{\zeta} \;-\; \pd{F_{0}}{\varepsilon} \pd{}{t} \right) \delta H 
\;\equiv\; \wh{{\cal F}}\;\delta H, 
\label{eq:deltaG_1} 
\end{eqnarray} 
where the perturbed Hamilton's equations for $\delta\dot{{\bf I}} = (\delta\dot{J}_{{\rm g}},\delta\dot{\varepsilon}, \delta\dot{J}_{{\rm d}})$ are given in 
Eqs.~(\ref{eq:delta_g})-(\ref{eq:delta_d}), and we have defined the operator $\wh{{\cal F}}$. 

\vspace*{0.2in} 

\no 
{\bf A. Slow-Time Evolution} 

\vspace*{0.2in} 

Since the guiding-center background distribution $F_{0}$ is {\it quasi-static} on the wave time scale and is independent of the azimuthal angle $\varphi$ and the gyro-angle $\zeta$, we introduce an averaging operation [denoted as $\ov{(\cdots)}$ and refered to as {\it wave} averaging] with respect to the fast wave-time-scale, the azimuthal angle, and the gyro-angle, with the property 
\begin{equation} 
\ov{F}_{0} = F_{0}({\bf I};\, \tau = \epsilon^{2}t). 
\end{equation} 
Applying the wave-averaging procedure on Eq.~(\ref{eq:Vlasov_start}), we obtain 
\begin{equation} 
\pd{F_{0}}{\tau} \;=\; -\; \ov{\left( \{ \delta F,\; \delta H \} \right)} \;=\; -\;\frac{1}{{\cal J}}\;\left\{\; \pd{}{I^{i}} \left[\; {\cal J} 
\ov{\left( \delta F\; \delta\dot{I}^{i} \right)}\;\right] \;+\; \;\pd{}{s} \left[\; {\cal J} \ov{\left( \delta F\;\delta\dot{s} \right)}\;\right] \;\right\}, 
\label{eq:F0_2} 
\end{equation} 
where the perturbed Vlasov distribution $\delta F$ and the perturbation Hamiltonian $\delta H$ have zero wave-averages: $\ov{\delta F} \;=\; 0 \;=\; \ov{\delta H}$ and expressions for $\delta\dot{I}^{i} = (\delta\dot{J}_{{\rm g}},\delta\dot{\varepsilon},\delta\dot{J}_{{\rm d}})$ and $\delta\dot{s}$ are found in 
Eqs.~(\ref{eq:delta_g})-(\ref{eq:delta_d}) and (\ref{eq:delta_s}), respectively. 

When the slow-time evolution equation (\ref{eq:F0_2}) for the guiding-center background distribution $F_{0}$ is expressed in terms of the non-adiabatic part $\delta G$ of the perturbed guiding-center distribution, we find (after some algebra) 
\begin{eqnarray} 
\pd{F_{0}}{\tau} & = & |v_{\|}| \left\{\; \pd{}{J_{{\rm d}}} \left[\;\frac{1}{|v_{\|}|}\ov{\left( \delta G\; \pd{\delta H}{\varphi}\right)}\;\right] \;+\; 
\pd{}{J_{{\rm g}}} \left[\;\frac{1}{|v_{\|}|}\ov{\left( \delta G\;\pd{\delta H}{\zeta} \right)} \;\right] \right. \nonumber \\ 
 &   &\mbox{}+\; \pd{}{\varepsilon} \left[\; \frac{\omega_{{\rm c}}}{|v_{\|}|}\ov{\left( \delta G\;\pd{\delta H}{\zeta} \right)} \;+\; \frac{\omega_{{\rm d}}}{|v_{\|}|} \ov{\left( \delta G\; \pd{\delta H}{\varphi} \right)} \;+\; \sigma \ov{\left( \delta G\;\pd{\delta H}{s} \right)} \;\right] \nonumber \\ 
 &  &\left.-\; \pd{}{s} \left[\; \frac{ac}{q|v_{\|}|}\ov{\left( \delta G\;\pd{\delta H}{\varphi}\right)} \;+\; \sigma\; \ov{\left( \delta G\; 
\pd{\delta H}{\varepsilon}\right)} \;-\; \frac{\sigma}{2}\; \frac{\partial^{2}F_{0}}{\partial\varepsilon^{2}} \;\ov{\left(\delta H^{2}\right)} \;\right] \;\right\}, 
\label{eq:F0_1} 
\end{eqnarray} 
where $\sigma = v_{\|}/|v_{\|}| = \pm\,1$ and we made use of the Liouville identities (\ref{eq:ids}) as well as the identities 
\[ \ov{(\delta F\,\partial_{\zeta}\delta H)} \;=\; \ov{(\delta G\,\partial_{\zeta}\delta H)} \;\;\;{\rm and}\;\;\; \ov{(\delta F\,\partial_{\varphi}\delta H)} \;=\; 
\ov{(\delta G\,\partial_{\varphi}\delta H)}. \] 
The slow-time evolution equation (\ref{eq:F0_1}) for the background distribution $F_{0}$ contains terms associated with exact derivatives in invariant ${\bf I}$-space (as expected) plus a term involving parallel spatial gradients $(\wh{{\sf b}}\bdot\nabla = \partial/\partial s$) along magnetic field lines. Since the background distribution $F_{0}$ is independent of the parallel spatial coordinate $s$ (i.e., $\partial F_{0}/\partial s = 0$), we must remove the parallel-gradient terms on the 
right side of Eq.~(\ref{eq:F0_1}) by introducing a second averaging operation. 

\vspace*{0.2in} 

\no 
{\bf B. Bounce Averaging and Fourier Decomposition} 

\vspace*{0.2in} 

To remove the parallel-gradient terms on the right side of Eq.~(\ref{eq:F0_1}), we introduce the bounce averaging operation 
\begin{equation} 
\langle\;\cdots\;\rangle \;=\; \frac{1}{\tau_{{\rm b}}}\;\sum_{\sigma}\;\int_{s_{L}}^{s_{U}}\; ds\;{\cal J} \;\left(\;\cdots\;\right), 
\label{eq:b_av} 
\end{equation} 
where ${\cal J} = 1/|v_{\|}|$ is the Jacobian introduced above, $s_{L}({\bf I})$ and $s_{U}({\bf I})$ are the turning points where the trapped particle's parallel velocity $v_{\|}$ vanishes, $\sum_{\sigma}$ denotes a sum over the two possible signs of $v_{\|} = \pm\,|v_{\|}|$, and $\tau_{{\rm b}}$ denotes the bounce period 
\[ \tau_{{\rm b}} \;=\; \sum_{\sigma}\;\int_{s_{L}}^{s_{U}}\; {\cal J}\;ds. \] 
The bounce-average operation defined here yields the following identities 
\[ \left\langle \frac{1}{{\cal J}}\;\pd{}{s} \left[\; {\cal J}\;\left(\;\cdots\;\right) \;\right] \right\rangle \;=\; 0 \;\;\;{\rm and}\;\;\; 
\left\langle \frac{1}{{\cal J}}\;\pd{}{I^{i}} \left[\; {\cal J}\;\left(\;\cdots\;\right) \;\right] \right\rangle \;=\; \frac{1}{\tau_{{\rm b}}}\; \left.\left.  \left. \left. \pd{}{I^{i}} \right[\;\tau_{{\rm b}}\;\right\langle (\cdots)\right\rangle \;\right]. \] 
Hence, by bounce averaging the slow-time evolution equation (\ref{eq:F0_1}) while using these identities, we obtain 
\begin{equation} 
\pd{F_{0}}{\tau} \;=\; -\;\frac{1}{\tau_{{\rm b}}}\; \pd{}{I^{i}} \left( \tau_{{\rm b}} \left\langle\; \ov{\delta G\;\delta\dot{I}^{i}} \;\right\rangle \right), 
\label{eq:F0_3} 
\end{equation} 
where we note that $\langle\ov{\delta\dot{I}^{i}}\rangle = 0$ and the bounce period $\tau_{{\rm b}}$ now appears as the new Jacobian. 

We now introduce the Fourier decomposition (assuming a discrete frequency spectrum $\{\omega_{k}\}$ for the waves) 
\begin{equation} 
\left( \begin{array}{c} 
\delta G \\ 
\\ 
\delta H 
\end{array} \right) \;=\; \sum_{k}\;\sum_{m,\ell = -\infty}^{\infty}\; 
\left( \begin{array}{c} 
\delta\wt{G}_{m\ell k}(s,\sigma;\,{\bf I}) \\ 
\\ 
\delta\wt{H}_{m \ell k}(s,\sigma;\,{\bf I}) 
\end{array} \right) \exp i\left( m\varphi + \ell\zeta - \omega_{k}t\right)  
\label{eq:Fourier_mlk} 
\end{equation} 
so that the fast-time evolution equation (\ref{eq:deltaG_1}) for the non-adiabatic part $\delta G$ can be written as 
\begin{equation} 
\left[\; \left. \left. v_{\|}\;\pd{}{s} \;-\; i\right( \omega_{k} \;-\; m\,\omega_{{\rm d}} \;-\; \ell\,\omega_{{\rm c}} \right) \;\right] \delta\wt{G}_{m\ell k} \;\equiv\; \wh{{\cal L}}\;\delta\wt{G}_{m\ell k} \;=\; i{\cal F}\; 
\delta\wt{H}_{m\ell k},
\label{eq:Lop_def}
\end{equation} 
where the differential operator $\wh{{\cal F}}$ [defined in Eq.~(\ref{eq:deltaG_1})] becomes 
\begin{equation} 
\wh{{\cal F}} \;\rightarrow\; i\,{\cal F} \;=\; i\; \left( m\;\pd{F_{0}}{J_{{\rm d}}} \;+\; \ell\;\pd{F_{0}}{J_{{\rm g}}} \;+\; \omega_{k}\;\pd{F_{0}}{\varepsilon} \right). 
\label{eq:F_op} 
\end{equation} 
Substituting the Fourier decomposition (\ref{eq:Fourier_mlk}) into Eq.~(\ref{eq:F0_3}), with Eqs.~(\ref{eq:delta_g}) and (\ref{eq:delta_d}), respectively, we find 
\begin{eqnarray} 
\left\langle \;\ov{\delta G\;\delta\dot{J}_{{\rm g}}}\;\right\rangle & = & -\;\sum_{m,\ell,k}\; \ell\; {\rm Im}\left\langle\delta\wt{G}_{m\ell k}\; 
\delta\wt{H}_{m\ell k}^{*} \right\rangle, \label{eq:QL_zeta} \\ 
 & & \nonumber \\ 
\left\langle \;\ov{\delta G\;\delta\dot{J}_{{\rm d}}}\;\right\rangle & = & -\;\sum_{m,\ell,k}\; m\;{\rm Im} \left\langle \delta\wt{G}_{m\ell 
k}\;\delta\wt{H}_{m\ell k}^{*} \right\rangle. \label{eq:QL_phi} 
\end{eqnarray} 
After performing various integrations by parts using Eqs.~(\ref{eq:delta_e}) and (\ref{eq:deltaG_1}), we also find 
\begin{eqnarray} 
\left\langle \;\ov{\delta G\;\delta\dot{\varepsilon}} \;\right\rangle & = & \left\langle \ov{\left[ \left( \frac{d_{0}\delta G}{dt} \;-\; \pd{\delta G}{t} \right) 
\delta H \right]}\right\rangle \;=\; \left\langle \ov{\left[ \left( \wh{{\cal F}}\,\delta H \;-\; \pd{\delta G}{t} \right) \delta H \right]} \right\rangle \nonumber \\ 
 & & \nonumber \\ 
& = & -\;\left\langle \ov{\left( \pd{\delta G}{t}\;\delta H \right)} \right\rangle \;=\; -\;\sum_{m,\ell,k}\; \omega_{k}\;{\rm Im}\left\langle\delta\wt{G}_{m\ell k} 
\;\delta\wt{H}_{m\ell k}^{*} \right\rangle, \label{eq:QL_energy} 
\end{eqnarray} 
where we used the identity $\langle\ov{(\wh{{\cal F}}\delta H)\,\delta H}\rangle = 0$. We note that Eqs.~(\ref{eq:QL_zeta})-(\ref{eq:QL_energy}) all involve sums 
containing the term ${\rm Im}\langle \delta\wt{G}_{m\ell k}\;\delta\wt{H}_{m\ell k}^{*}\rangle$, so following our earlier work (Paper I) we introduce the 
{\it quasilinear potential} 
\begin{equation} 
\Gamma_{m\ell k} \;\equiv\; {\cal F}^{-1}\;{\rm Im}\left\langle\delta\wt{G}_{m\ell k}\; \delta\wt{H}_{m\ell k}^{*} \right\rangle, 
\label{eq:QL_Pot} 
\end{equation} 
so that we may replace 
\begin{equation} 
{\rm Im}\left\langle\delta\wt{G}_{m\ell k}\;\delta\wt{H}_{m\ell k}^{*} \right\rangle \;=\; \Gamma_{m\ell k} \; \left( \ell\;\pd{F_{0}}{J_{{\rm g}}} \;+\; 
\omega_{k}\;\pd{F_{0}}{\varepsilon} \;+\; m\;\pd{F_{0}}{J_{{\rm d}}} \right) 
\label{eq:Gamma_F} 
\end{equation} 
into Eqs.~(\ref{eq:QL_zeta})-(\ref{eq:QL_energy}). 

\vspace*{0.2in} 

\no 
{\bf C. Relativistic Quasilinear Diffusion Tensor} 

\vspace*{0.2in} 

By substituting Eqs.~(\ref{eq:QL_zeta})-(\ref{eq:QL_energy}) and (\ref{eq:Gamma_F}) into Eq.~(\ref{eq:F0_3}), we obtain the relativistic quasilinear diffusion equation \begin{equation} 
\pd{F_{0}({\bf I},\tau)}{\tau} \;=\; \frac{1}{\tau_{{\rm b}}} \pd{}{I^{i}}\; \left( \tau_{{\rm b}}\;D^{ij}_{{\rm QL}}\; \pd{F_{0}({\bf I},\tau)}{I^{j}} 
\right), \label{eq:RQLE} 
\end{equation} 
where the invariant coordinates are the gyro-action $I^{1} = J_{{\rm g}}$, the particle guiding-center kinetic energy $I^{2} = \varepsilon$, and the drift action 
$I^{3} = J_{{\rm d}}$. In Eq.~(\ref{eq:RQLE}), the relativistic quasilinear diffusion tensor ${\sf D}_{{\rm QL}}$ has the following symmetric form 
\begin{equation} 
{\sf D}_{{\rm QL}} \;=\; \sum_{m,\ell,k}\; \left( \begin{array}{ccc} 
\ell^{2} & \ell\omega_{k} & \ell m \\ 
\omega_{k}\ell & \omega_{k}^{2} & \omega_{k}m \\ 
m \ell & m\omega_{k} & m^{2} 
\end{array} \right)\;\Gamma_{m\ell k}. 
\label{eq:DRQL} 
\end{equation} 
Note the simplicity of the quasilinear coefficient matrix ${\sf d}_{{\rm m}k}$ when expressed in terms of the invariant coordinates ${\bf I} = (J_{{\rm g}},\varepsilon,
J_{{\rm d}})$. Since the quasilinear transport of trapped particles (e.g., see Ref.~\cite{schulz74}) is often discussed in terms of the equatorial pitch angle instead of the gyro-action $J_{{\rm g}}$, an alternative representation of the quasilinear coefficient matrix expressed in terms of the equatorial pitch-angle instead of the gyro-action is presented in Appendix A. Although the pitch-angle formulation will facilitate comparison with earlier works, however, the simplicity of the 
quasilinear coefficient matrix in Eq.~(\ref{eq:DRQL}) is lost. 

\vspace*{0.2in} 

\no 
{\sf IV. QUASILINEAR POTENTIAL} 

\vspace*{0.2in} 

To complete the derivation of the relativistic quasilinear diffusion tensor ${\sf D}_{{\rm QL}}$, with components given by Eq.~(\ref{eq:DRQL}), we must now solve for an explicit expression for the quasilinear potential $\Gamma_{m\ell k}$. 

\vspace*{0.2in} 

\no 
{\bf A. Guiding-Center Perturbation Hamiltonian} 

\vspace*{0.2in} 

In the present work (as in Paper I \cite{BC_2001}), we use the gauge condition $\delta{\bf A}\bdot\nabla\varphi = 0$ so that the perturbed vector potential is written as 
\begin{equation} 
\delta{\bf A} \;=\; \delta A_{\|}\;\nabla s \;-\; \delta\beta\;\nabla\psi, 
\end{equation} 
where $\delta A_{\|} = \partial\delta\alpha/\partial s$ denotes the parallel component of the perturbed vector potential and the perpendicular components of the perturbed vector potential $\delta{\bf A}_{\bot} = \delta{\bf A} - \delta A_{\|}\; \wh{{\sf b}}$ are 
\begin{equation} 
\delta{\bf A}_{\bot} \;=\; -\; \left( \delta\beta \;+\; a\;\pd{}{s}\,\delta\alpha \right) \nabla\psi. 
\label{eq:deltaA_perp} 
\end{equation} 
Hence, the parallel component of the perturbed magnetic field is 
\begin{equation} 
\delta B_{\|} \;=\; B\;\pd{}{\varphi} \left( \delta\beta \;+\; a\;\pd{\delta\alpha}{s} \right), 
\end{equation} 
while the parallel component of the perturbed electric field is 
\begin{equation} 
\delta E_{\|} \;=\; -\;\wh{{\sf b}}\bdot \left( \nabla\delta\phi \;+\; \frac{1}{c}\;\pd{\delta{\bf A}}{t} \right) \;=\; -\;\pd{\delta\Phi}{s}, 
\end{equation} 
where $\delta\Phi = \delta\phi + c^{-1}\partial_{t}\delta\alpha$ denotes the effective perturbation scalar potential. Note that the Fourier components of the perturbed parallel electric and magnetic fields are given as 
\begin{eqnarray} 
\delta E_{\| mk} & = & -\;\pd{\delta\Phi_{mk}}{s} \;=\; -\;\pd{}{s} \left( \delta\phi_{mk} \;-\; i\;\frac{\omega_{k}}{c}\;\delta\alpha_{mk} \right), 
\label{eq:E_mlk} \\ \delta B_{\| mk} & = & i\;mB\; \left( \delta\beta_{mk} \;+\; a\;\pd{\delta\alpha_{mk}}{s} \right). \label{eq:B_mlk} 
\end{eqnarray} 
As in Paper I \cite{BC_2001}, it is possible to find expressions for the perturbation potentials $\delta\phi_{mk}$, $\delta\alpha_{mk}$, and $\delta\beta_{mk}$ in terms of electric covariant components $\delta E_{i\;mk} = \delta{\bf E}_{mk}\bdot\partial_{i}{\bf X}$ and magnetic contravariant components $\delta B_{mk}^{i} = 
\delta{\bf B}_{mk}\bdot\nabla\psi^{i}$. 

\vspace*{0.1in} 

\no 
{\bf B. Fourier-Bessel Expansions} 

\vspace*{0.2in} 

Using the gauge condition introduced in the previous Section, the perturbation Hamiltonian (\ref{eq:delta_Hgc}) becomes 
\begin{equation} 
\delta H \;=\; q \left( \delta_{gc}\phi \;-\; \frac{v_{\|}}{c}\;\pd{}{s}\,\delta_{gc}\alpha \right) \;+\; \frac{q}{c}\;{\bf v}_{\bot}\bdot\nabla\psi \left( 
\delta_{gc}\beta \;+\; a\;\pd{}{s}\,\delta_{gc}\alpha \right). 
\label{eq:deltaH_gauge} 
\end{equation} 
We now consider the perturbation potential 
\[ \delta\chi({\bf X} + \vb{\rho},t) \;=\; \sum_{k}\; e^{-i\omega_{k}t}\;\delta\chi_{k}({\bf X} + \vb{\rho}), \] 
which, using Eqs.~(\ref{eq:chi_gc})-(\ref{eq:Taylor_chi}), yields the azimuthal-angle Fourier expansion 
\begin{equation} 
\delta\chi_{k}({\bf X} + \vb{\rho}) \;=\; \sum_{m = -\infty}^{\infty}\;e^{im\,\varphi} \left[\; \exp\left(-i\,\eta\,\sin\zeta \;+\; \cos\zeta\;\wh{\lambda}\right)\;\delta\chi_{mk}(\psi,\,s) \;\right], 
\label{eq:Fourier_phi} 
\end{equation} 
where 
\begin{equation} 
\eta \;\equiv\; m\;\frac{\rho}{R} \;\;\;{\rm and}\;\;\; \wh{\lambda} \;\equiv\; \Delta\psi \left( \pd{}{\psi} \;-\; a\;\pd{}{s} \right). 
\label{eq:eta} 
\end{equation} 
We note that the argument $\wh{\lambda}$ is a differential operator acting only on the perturbation fields. 

Next, we introduce the following Bessel-function identities 
\begin{eqnarray} 
\exp\left(-\,i\;\eta\;\sin\zeta\right) & = & \sum_{\jmath = -\infty}^{\infty}\; \exp(-\,i\jmath\zeta)\;J_{\jmath}(\eta) \\ 
\exp\left(\cos\zeta\;\wh{\lambda}\right) & = & \sum_{\jmath^{\prime} = -\infty}^{\infty}\; \exp(i\jmath^{\prime}\zeta)\;I_{\jmath^{\prime}}(\wh{\lambda}), 
\end{eqnarray} 
where $J_{\jmath}$ and $I_{\jmath^{\prime}}$ denote the Bessel and modified Bessel functions of order $\jmath$ and $\jmath^{\prime}$, respectively, and 
\begin{equation} 
\frac{1}{2\pi} \oint\; e^{-\,i\ell\zeta} \left( \sum_{\jmath = -\infty}^{\infty}\; e^{-\,i\jmath\zeta}\;J_{\jmath}(\eta) \right) \left( \sum_{\jmath^{\prime} = 
-\infty}^{\infty}\;e^{i\jmath^{\prime}\zeta}\;I_{\jmath^{\prime}}(\wh{\lambda}) \right) \; d\zeta \;=\; \sum_{\jmath = -\infty}^{\infty}\; J_{\jmath}\;I_{\jmath + \ell}. 
\end{equation} 
The Fourier gyroangle expansion of the generic perturbed potential (\ref{eq:Fourier_phi}) is, therefore, expressed as 
\begin{eqnarray} 
\delta\chi_{k}({\bf X} + \vb{\rho}) & = & \sum_{m,\,\ell = -\infty}^{\infty}\; e^{i\,(m\varphi + \ell\zeta)}\; \left( \sum_{\jmath = -\infty}^{\infty}\;
J_{\jmath}(\eta)\;I_{\jmath + \ell}(\wh{\lambda}) \right)\;\delta\chi_{mk}(\psi,s) \nonumber \\ 
& = & \sum_{m,\,\ell =  -\infty}^{\infty}\; e^{i\,(m\varphi + \ell\zeta)}\; \delta\wt{\chi}_{m\ell  k}, \label{eq:chi_mlk} 
\end{eqnarray} 
where the Fourier-Bessel components $\delta\wt{\chi}_{m\ell k}(s;\;{\bf I})$ are functions of the parallel spatial coordinate $s$ and the invariant 
coordinates ${\bf I} = (J_{{\rm g}},\varepsilon,J_{{\rm d}})$. 

Lastly, using the expression (\ref{eq:B_mlk}) for $\delta\wt{B}_{\| m\ell k}$, the Fourier-Bessel expansion of the third term ${\bf v}_{\bot}\bdot\delta{\bf A}_{\bot}$ in the perturbation Hamiltonian (\ref{eq:deltaH_gauge}), with $\delta{\bf A}_{\bot}$ given by Eq.~(\ref{eq:deltaA_perp}), is progressively transformed as follows. First, 
we begin with 
\begin{eqnarray*} 
-\;\frac{q}{c}\;{\bf v}_{\bot}\bdot\delta{\bf A}_{\bot k} & = & -\;\frac{q}{c}\;|\nabla\psi|\;v_{\bot}\sin\zeta\; \left( \delta\beta_{k} \;+\; a\;\pd{}{s}\,
\delta\alpha_{k} \right) \\ 
 & = & i\;\omega_{{\rm c}}Rp_{\bot}\;\sin\zeta \;\sum_{m, \ell}\; \frac{e^{i(m\varphi + \ell\zeta)}}{m} \left( \sum_{\jmath}\; J_{\jmath}\;I_{\jmath + \ell} \right) \frac{\delta B_{\| mk}}{B}, 
\end{eqnarray*} 
where Eqs.~(\ref{eq:B_mlk}) and (\ref{eq:chi_mlk}) were used, with the definitions $|\nabla\psi| = BR$ and $p_{\bot} = \gamma\,Mv_{\bot}$. Next, by 
substituting $\sin\zeta = (e^{i\zeta} - e^{-i\zeta})/2i$, we obtain 
\[ -\;\frac{q}{c}\;{\bf v}_{\bot}\bdot\delta{\bf A}_{\bot k} \;=\; \omega_{{\rm c}}Rp_{\bot} \;\sum_{m, \ell}\; \frac{e^{i(m\varphi + \ell\zeta)}}{2\,m} \;\left[\; 
\left. \left. \sum_{\jmath}\; I_{\jmath + \ell} \right( J_{\jmath + 1} \;-\; J_{\jmath - 1} \right) \;\right] \frac{\delta B_{\| mk}}{B}, \] 
where the $\jmath$-summation was re-arranged. Lastly, we use the Bessel recurrence relation $J_{\jmath - 1}(\eta) - J_{\jmath + 1}(\eta) = 2\eta^{-1}\,
J^{\prime}_{\jmath}(\eta)$ to obtain 
\begin{eqnarray} 
-\;\frac{q}{c}\;{\bf v}_{\bot}\bdot\delta{\bf A}_{\bot k} & = & J_{{\rm g}}\omega_{{\rm c}} \;\sum_{m, \ell}\; e^{i(m\varphi + \ell\zeta)} \;\left[\; 
\sum_{\jmath}\; I_{\jmath + \ell}(\wh{\lambda}) \;\left( -\;\frac{2}{\eta}\;J^{\prime}_{\jmath}(\eta) \right) \;\right] \frac{\delta B_{\| mk}}{B} \nonumber \\ & = & 
J_{{\rm g}}\omega_{{\rm c}}\; \sum_{m,\,\ell = -\,\infty}^{\infty}\; e^{i(m\varphi + \ell\zeta)} \;\frac{\delta \wt{B}_{\| m\ell k}}{B}. 
\end{eqnarray} 
In summary, the Fourier-Bessel components of the perturbation Hamiltonian (\ref{eq:deltaH_gauge}) are 
\begin{equation} 
\delta \wt{H}_{m\ell k} \;=\; q \left( \delta\wt{\phi}_{m\ell k} \;-\; \sigma\;\frac{|v_{\|}|}{c}\; \pd{\delta\wt{\alpha}_{m\ell k}}{s} \right) \;+\; J_{{\rm g}}
\omega_{{\rm c}}\; \frac{\delta \wt{B}_{\| m\ell k}}{B}, 
\label{eq:H_mlk} 
\end{equation} 
where 
\begin{eqnarray} 
\left( 
\begin{array}{c} 
\delta\wt{\phi}_{m\ell k} \\ 
\\ 
\delta\wt{\alpha}_{m\ell k} 
\end{array} \right) & = & \sum_{\jmath = -\infty}^{\infty}\; J_{\jmath}(\eta)\;I_{\jmath + \ell}(\wh{\lambda})\; \left(  
\begin{array}{c} 
\delta\phi_{mk}(\psi,s) \\ 
\\ 
\delta \alpha_{mk}(\psi,s) \end{array} 
\right) \\ \nonumber 
\\ 
\delta\wt{B}_{\| m\ell k} & = & \sum_{\jmath = -\infty}^{\infty}\;\left( -\;\frac{2}{\eta}\;J^{\prime}_{\jmath}(\eta) \right)\; I_{\jmath + \ell}(\wh{\lambda}) \;
\delta B_{\| mk}(\psi,s). 
\end{eqnarray} 
The Fourier-Bessel components $\delta \wt{H}_{m\ell k}(s,\sigma;\,{\bf I})$ of the perturbation guiding-center Hamiltonian can now be used explicitly in the expression for the quasilinear potential (\ref{eq:QL_Pot}). 

\vspace*{0.2in} 

\no 
{\bf C. Solution of Fast-Time Evolution Equation} 

\vspace*{0.2in} 

Following an approach detailed in Paper I, we remove the $\sigma$-dependence of $\delta\wt{H}_{m\ell k}(s,\sigma;\,{\bf I})$ by introducing a new perturbation Hamiltonian 
\begin{eqnarray} 
\delta\wt{K}_{m\ell k} & \equiv & \delta\wt{H}_{m\ell k} \;+\; \frac{q}{c}\;\wh{{\cal L}}\;\delta\wt{\alpha}_{m\ell k} \nonumber \\
  & = & q\,\delta\wt{\Phi}_{m\ell k} \;+\; \left. \left. i\,\frac{q}{c}\;\right( m\,\omega_{{\rm d}} \;+\; \ell\,\omega_{{\rm c}} \right)\;\delta\wt{\alpha}_{m\ell k} 
\;+\; J_{{\rm g}}\omega_{{\rm c}}\; \frac{\delta \wt{B}_{\| m\ell k}}{B} 
\label{eq:ndkmlk} 
\end{eqnarray} 
where the differential operator $\wh{{\cal L}}$ is defined in Eq.~(\ref{eq:Lop_def}) and 
\[ \delta\wt{\Phi}_{m\ell k} \;=\; \delta\wt{\phi}_{m\ell k} \;-\; i\,(\omega_{k}/c)\,\delta\wt{\alpha}_{m\ell k} \;=\; -\; \int\;\delta\wt{E}_{\| m\ell k}\;ds. \] 
Note that, as in Paper I (Appendix B), the new perturbation Hamiltonian can also be expressed as 
\begin{equation} 
\delta\wt{K}_{m{\ell}k} \;=\; \frac{q}{mc} (m\omega_{{\rm d}} + \ell\omega_{{\rm c}} - \omega_{k}) \,\int \frac{\delta\wt{B}_{m\ell k}^{\psi}}{B} ds \;+\; J_{{\rm g}}\omega_{{\rm c}}\, \frac{\delta\wt{B}_{\parallel\,m\ell{k}}}{B} \;+\; \frac{iq}{m} \delta\wt{E}_{\varphi\,m\ell{k}}, \label{eq:dkmlk} 
\end{equation} 
in terms of components of the perturbed electric and magnetic fields. 

We also introduce the new nonadiabatic part $\delta\wt{G}^{\prime}_{m\ell k}$ defined as 
\begin{equation} 
\delta\wt{G}^{\prime}_{m\ell k} \;=\; \delta\wt{G}_{m\ell k} \;+\; i\;\frac{q}{c}\;{\cal F}\,\delta\wt{\alpha}_{m\ell k} 
\end{equation} 
so that the fast time scale evolution for $\delta\wt{G}^{\prime}_{m\ell k}$ is obtained from Eq.~(\ref{eq:deltaG_1}) as 
\begin{eqnarray} 
\wh{{\cal L}}\;\delta\wt{G}^{\prime}_{m\ell k} & = & \wh{{\cal L}}\;\delta\wt{G}_{m\ell k} \;+\; i\,\frac{q}{c}\; {\cal F}\;\wh{{\cal L}}\;\delta\wt{\alpha}_{m\ell k} \nonumber \\  
& = & i\,{\cal F} \left( \delta\wt{H}_{m\ell k} \;+\; i\,\frac{q}{c}\;\wh{{\cal L}}\;\delta\wt{\alpha}_{m\ell k} \right) \;=\; i\;{\cal F}\,\delta\wt{K}_{m\ell k}, \label{eq:deltaG_prime} 
\end{eqnarray} 
where we used the fact that the differential operator $\wh{{\cal L}}$ commutes with ${\cal F}$. 

The rest of the analysis leading to the solution of $\delta\wt{G}_{m\ell k}^{\prime}$ in terms of the modified perturbation Hamiltonian $\delta\wt{K}_{m\ell k}$ follows Paper I \cite{BC_2001}. First, we note that the transformation $(\delta\wt{G}_{m\ell k},\delta\wt{H}_{m\ell k}) \rightarrow (\delta\wt{G}^{\prime}_{m\ell k}, 
\delta\wt{K}_{m\ell k})$ leaves the quasilinear potential invariant: 
\begin{equation} 
\Gamma_{m\ell k} \;=\; {\cal F}^{-1}\;{\rm Im} \left\langle \delta\wt{G}^{\prime}_{m\ell k}\; \delta\wt{K}_{m\ell k}^{*}\right\rangle, 
\label{eq:Gamma_prime} 
\end{equation} 
as follows from properties of the averaging operations $\langle \ov{(\cdots)}\rangle$ and the differential operator $\wh{{\cal L}}$. From Eq.~(61) of Paper I, where we now replace $m\,\omega_{{\rm d}} \rightarrow (m\,\omega_{{\rm d}} + \ell\,\omega_{{\rm c}})$, the solution of Eq.~(\ref{eq:deltaG_prime}) is written as 
\begin{eqnarray} 
\delta\wt{G}^{\prime}_{m\ell k} & = & {\cal F}\,e^{i\sigma\theta} \left\{\; i\sigma\;\int_{s_{L}}^{s}\; \frac{ds^{\prime}}{|v_{\|}|}\; 
e^{-i\,\sigma\theta(s^{\prime})}\;\delta\wt{K}_{m\ell k}(s^{\prime}) \right. \nonumber \\  
&  &\hspace*{0.5in}\left.-\; \left. \left. \frac{\tau_{{\rm b}}}{2} \right( \cot\Theta \left\langle \delta\wt{K}_{m\ell k}\;\cos\theta \right\rangle \;+\; 
\left\langle \delta\wt{K}_{m\ell k}\;\sin\theta \right\rangle \right) \;\right\}, \label{eq:G_sol} 
\end{eqnarray} 
where the $s$-dependent angle $\theta(s)$ is defined as 
\begin{equation} 
\theta(s;\,{\bf I}) \;=\; \left. \left. \int_{s_{L}}^{s}\; \frac{ds^{\prime}}{|v_{\|}|}\; \right( \omega_{k} \;-\; m\, \omega_{{\rm d}}(s^{\prime}) \;-\; 
\ell\,\omega_{{\rm c}}(s^{\prime}) \right), 
\label{eq:thetas} 
\end{equation} 
while the $s$-independent angle $\Theta$ is defined as 
\begin{equation} 
\Theta({\bf I}) \;=\; \left. \left. \frac{\tau_{{\rm b}}}{2} \right( \omega_{k} \;-\; \ell\,\langle\omega_{{\rm c}}\rangle \;-\; m\,\langle\omega_{{\rm d}}\rangle 
\right). \label{eq:Theta_av} 
\end{equation} 
Here, $\langle\omega_{{\rm c}}\rangle$ and $\langle\omega_{{\rm d}}\rangle$ denote the bounce-averages of the gyro-frequency and drift-frequency, respectively; in 
Appendix B, we present a simple expression for the bounce-averaged drift frequency $\langle\omega_{{\rm d}}\rangle$ in terms of the bounce-averaged radial gradient 
$\langle\partial\omega_{{\rm c}}/\partial J_{{\rm d}}\rangle$. 

Using simple relations derived in Paper I, we ultimately find the quasilinear potential $\Gamma_{m\ell k}$ given as 
\begin{equation} 
\Gamma_{m\ell k}({\bf I}) \;=\; \frac{\tau_{{\rm b}}}{2}\; \left|\left\langle \delta\wt{K}_{m\ell k}(s;\,{\bf I})\; \cos\theta(s;\,{\bf I})\right\rangle\right|^{2}\; 
{\rm Im}(-\,\cot\Theta). 
\label{eq:Gamma_cot} 
\end{equation} 
Here, we note that 
\begin{equation} 
\frac{\tau_{{\rm b}}}{2}\;\cot\Theta \;=\; \left. \left. \sum_{n = -\infty}^{\infty}\; \right( \omega_{k} - \ell\,\langle \omega_{{\rm c}}\rangle - m\, 
\langle\omega_{{\rm d}}\rangle - n\,\omega_{{\rm b}}\right)^{-1}, 
\end{equation} 
where $\omega_{{\rm b}} = 2\pi/\tau_{{\rm b}}$ denotes the bounce frequency. 

Lastly, using the Plemelj formula, we find 
\begin{equation} 
\frac{\tau_{{\rm b}}}{2}\;{\rm Im}(-\,\cot\Theta) \;=\; \sum_{n = -\infty}^{\infty}\; \pi\;\delta\left( \omega_{k} \;-\; {\bf m}\bdot\langle\vb{\omega}_{{\bf m}}
\rangle  \right) 
\end{equation} 
where ${\bf m} = (\ell,n,m)$ and 
\begin{equation} 
{\bf m}\bdot\langle\vb{\omega}_{{\bf m}}\rangle \;\equiv\; \ell\,\langle\omega_{{\rm c}}\rangle + n\,\omega_{{\rm b}} + m\,\langle\omega_{{\rm d}}\rangle. 
\end{equation} 
Using these expressions, the quasilinear potential (\ref{eq:Gamma_cot}) is written as 
\begin{equation} 
\Gamma_{m\ell k} \;=\; \sum_{n = -\infty}^{\infty}\; \pi\;\delta\left( \omega_{k} - {\bf m}\bdot\langle\vb{\omega}_{{\bf m}}\rangle \right) 
\left|\left\langle \delta\wt{K}_{m\ell k}\;\cos\theta\right\rangle\right|^{2}. 
\label{eq:Gamma_mlk} 
\end{equation} 
Note that wave-particle resonances involve harmonics of the bounce-averaged gyrofrequency $\langle\omega_{{\rm c}}
\rangle$ and the bounce-averaged drift-frequency $\langle\omega_{{\rm d}}\rangle$. By combining these results into 
Eq.~(\ref{eq:DRQL}), we finally obtain the expression for the relativistic quasilinear diffusion tensor 
\begin{equation} 
{\sf D}_{{\rm QL}} \;=\; \sum_{{\bf m},k}\; \left( \begin{array}{ccc} 
\ell^{2} & \ell\omega_{k} & \ell m \\ 
\omega_{k}\ell & \omega_{k}^{2} & \omega_{k}m \\ 
m \ell & m\omega_{k} & m^{2} 
\end{array} \right)\; \left[\; \pi\;\delta\left( \omega_{k} - {\bf m}\bdot 
\langle\vb{\omega}_{{\bf m}}\rangle \right) \left|\left\langle 
\delta\wt{K}_{m\ell k} \;\cos\theta\right\rangle\right|^{2} \;\right]. \label{eq:DRQL_mlk} 
\end{equation} 
The structure of the relativistic quasilinear diffusion tensor (\ref{eq:DRQL_mlk}) clearly shows that, for arbitrary values of gyro-harmonic and drift-harmonic numbers $(\ell, m)$ and non-vanishing wave frequency $\omega_{k}$, off-diagonal quasilinear transport cannot be neglected (see Appendix A for a brief discussion of off-diagonal quasilinear transport coefficients). Note also that wave-particle resonances, in fact, involve bounce-averaged cyclotron and drift frequencies, not their local expressions as might be expected. 
  
\vspace*{0.2in} 

\no 
{\sf V. SUMMARY} 

\vspace*{0.2in} 

The present work has presented a complete derivation of the relativistic bounce-averaged quasilinear diffusion equation associated with arbitrary-frequency electromagnetic fluctuations in axisymmetric geometry. The main results are the relativistic quasilinear equation, Eq.~(\ref{eq:RQLE}) and the corresponding diffusion tensor, Eq.~(\ref{eq:DRQL_mlk}), where $\delta\wt{K}_{m{\ell}k}$ is given by Eq.~(\ref{eq:ndkmlk}) or Eq.~(\ref{eq:dkmlk}). In Eq.~(\ref{eq:DRQL_mlk}) the $s$-dependent angle $\theta(s)$ is given by Eq.~(\ref{eq:thetas}), $\langle...\rangle$ denotes bounce-averaging, and the Fourier-Bessel components (denoted by a tilde) are defined in Eq.~(\ref{eq:chi_mlk}). The quasilinear diffusion tensor is also presented in energy, pitch-angle coordinates in Appendix A. 

Future work will include a comparative study with previous analytical models of pitch-angle and energy diffusion 
\cite{lyons71,lyons74}, and with calculations of radial diffusion coefficients \cite{schulz74,elkington99,elkington03}, plus generalization to allow non-axisymmetric unperturbed electric and magnetic fields. In related work, a numerical implementation of the present relativistic quasilinear diffusion equation in a multi-dimensional simulation code is being developed (in collaboration with Dr. Jay Albert), including an investigation of the importance of off-diagonal quasilinear diffusion coefficients in non-axisymmetric magnetic geometries. 

\vspace*{0.2in} 

\no 
{\sf ACKNOWLEDGEMENTS} 

\vspace*{0.2in} 

The authors thank Dr. Jay Albert and Professor Liu Chen for helpful discussions. This work was supported by the National Science Foundation under grant number ATM-0316195 and under the Boston University NSF Center for Integrated Space Weather Modeling (CISM, contract number GC177024NGA), and by the NASA Sun-Earth Connections Theory Program under NASA grant number NAG5-11881. 

\vspace*{0.2in} 

\setcounter{equation}{0} 
\renewcommand{\theequation}{A.\arabic{equation}} 

\no 
{\bf Appendix A. Equatorial Pitch-Angle Formulation} 

\vspace*{0.2in} 

Since quasilinear transport is often discussed in terms of the equatorial pitch-angle instead of the gyro-action 
$J_{{\rm g}}$ \cite{schulz74}, we introduce the cosine of the equatorial pitch angle (denoted $\xi$) defined as 
\begin{equation} 
\xi(J_{{\rm g}},\varepsilon,J_{{\rm d}}) \;=\; \frac{p_{\| 0}}{p} \;=\; \left(\; 1 \;-\; \frac{J_{{\rm g}} 
\omega_{{\rm g}0}}{\varepsilon\,(\gamma + 1)/2}  \;\right)^{\frac{1}{2}} \label{eq:pitch} 
\end{equation} 
where the parallel momentum $p_{\| 0}$ and the rest-mass gyro-frequency $\omega_{{\rm g}0} = qB_{0}(\psi)/Mc$ are evaluated on the equatorial plane. From this definition, we obtain the differential relation 
\begin{equation} 
d\xi \;=\; \frac{1}{J_{\xi}\omega_{{\rm c}0}} \left[\; \omega_{{\rm c}0}\;dJ_{{\rm g}} \;+\; \omega_{{\rm d}0}\;
dJ_{{\rm d}} \;-\; \left( 1 - \xi^{2} \right) d\varepsilon \;\right], 
\label{eq:xi_diff} 
\end{equation} 
where the drift frequency 
\begin{equation} 
\omega_{{\rm d}0} \;\equiv\; J_{{\rm g}}\,\pd{\omega_{{\rm c}0}}{J_{{\rm d}}} \label{eq:omegad_0} 
\end{equation} 
defines an equatorial drift frequency [see Eq.~(\ref{eq:omegad_average})] and 
\begin{equation} 
J_{\xi}(\xi,\varepsilon, J_{{\rm d}}) \;=\; \frac{pp_{\| 0}}{M\omega_{{\rm g}0}} \label{eq:Jac_xi} 
\end{equation} 
is the Jacobian associated with the substitution $J_{{\rm g}} \rightarrow \xi$ (and thus $J_{\xi}$ has units of action). Using the differential relation (\ref{eq:xi_diff}), the unperturbed guiding-center evolution of the equatorial pitch angle is $\dot{\xi}_{0} \;=\; 0$. 

The perturbed pitch-angle Hamilton's equation for $\delta\dot{\xi}$ is expressed as 
\[ \delta\dot{\xi} \;=\; \frac{-1}{J_{\xi}\omega_{{\rm c}0}} \left[\; \omega_{{\rm c}0}\;\pd{}{\zeta} \;+\; 
\omega_{{\rm d}0}\;\pd{}{\varphi} \;+\; \left(1 - \xi^{2}\right) \left( \pd{}{t} - \frac{d_{0}}{dt}\right) \;\right] \delta H. \] 
The differential operator $\wh{{\cal F}}$ becomes 
\begin{equation} 
\wh{{\cal F}} \;=\; \left( \pd{F_{0}}{J_{{\rm d}}} \;+\; \frac{\omega_{{\rm d}0}}{J_{\xi}\omega_{{\rm c}0}}\; \pd{F_{0}}{\xi} \right) \pd{}{\varphi} \;+\; \frac{1}{J_{\xi}}\;\pd{F_{0}}{\xi}\;\pd{}{\zeta} \;-\; \left( \pd{F_{0}}{\varepsilon} 
\;-\; \frac{(1 - \xi^{2})}{J_{\xi}\omega_{{\rm c}0}}\;\pd{F_{0}}{\xi}\right) \pd{}{t} 
\label{eq:F_xi} 
\end{equation} 
Relativistic quasilinear diffusion equation (\ref{eq:RQLE}) can also be written in terms of the invariant coordinates 
$\ov{{\bf I}} = (\xi, \varepsilon, J_{{\rm d}})$ 
\begin{equation} 
\pd{F_{0}}{\tau} \;=\; \frac{1}{J_{\xi}\tau_{{\rm b}}} \pd{}{\ov{I}^{i}}\; \left( J_{\xi}\tau_{{\rm b}}\; \ov{D}^{ij}_{{\rm QL}}\;\pd{F_{0}}{\ov{I}^{j}} \right) 
\end{equation} 
where $J_{\xi}$ is defined in Eq.~(\ref{eq:Jac_xi}) and the components of the new quasilinear diffusion tensor are defined by the relation
\begin{equation} 
\ov{D}^{ij}_{{\rm QL}} \;=\; \pd{\ov{I}^{i}}{I^{a}}\;D^{ab}_{{\rm QL}}\;\pd{\ov{I}^{j}}{I^{b}}. \label{eq:DQL_pitch} 
\end{equation} 
The new quasilinear diffusion tensor is, therefore, expressed as
\begin{equation} 
\ov{{\sf D}}_{{\rm QL}} \;=\; \sum_{m,\ell,k}\; \left( \begin{array}{ccc} 
\Lambda_{m\ell k} & \omega_{k}\,\lambda_{m\ell k} & m\,\lambda_{m\ell k} \\ 
\omega_{k}\,\lambda_{m\ell k} & \omega_{k}^{2} & \omega_{k}m \\ 
m\,\lambda_{m\ell k} & m\omega_{k} & m^{2} 
\end{array} \right)\;\Gamma_{m\ell k},
\label{eq:DRQL_pitch} 
\end{equation} 
where the pitch-angle coefficients $\Lambda_{m\ell k}$ and $\lambda_{m\ell k}$ are defined as
\begin{eqnarray}
\Lambda_{m\ell k} & = & \frac{1}{J_{\xi}^{2}\omega_{{\rm c}0}^{2}} \left[\; \ell^{2}\omega_{{\rm c}0}^{2} \;+\; 
(1 - \xi^{2})^{2}\;\omega_{k}^{2} \;+\; m^{2}\,\omega_{{\rm d}0}^{2} \;\right], \label{eq:Pitch_diag} \\
 &  & \nonumber \\
\lambda_{m\ell k} & = & \frac{1}{J_{\xi}\omega_{{\rm c}0}} \left[\; \ell\,\omega_{{\rm c}0} \;-\; 
(1 - \xi^{2})\;\omega_{k} \;+\; m\,\omega_{{\rm d}0} \;\right]. \label{eq:Pitch_off}
\end{eqnarray}
Note that, while the diagonal coefficient (\ref{eq:Pitch_diag}) cannot vanish (since it is strictly positive), the off-diagonal coefficient (\ref{eq:Pitch_off}) may be small when evaluated at the wave-particle resonance $\omega_{k} = \ell\langle \omega_{{\rm c}}\rangle + n\,\omega_{{\rm b}} + m\,\langle\omega_{{\rm d}}\rangle$. Further discussion of the importance of these off-diagonal quasilinear transport coefficients is, however, outside the scope of the present work and will be a subject for future work when non-axisymmetric magnetic geometries are also considered.

\vspace*{0.2in} 

\setcounter{equation}{0} \renewcommand{\theequation}{B.\arabic{equation}} 

\no 
{\bf Appendix B. Bounce-Averaged Drift Frequency} 

\vspace*{0.2in} 

In this Appendix, we derive an explicit expression for the bounce-averaged drift frequency based on the definition (\ref{eq:axi_drift}): 
\begin{equation} 
\langle \omega_{{\rm d}}\rangle \;=\; \frac{1}{\tau_{{\rm b}}} \sum_{\sigma}\;\int ds\; \left( \frac{J_{{\rm g}}}{|v_{\|}|}\;\pd{\omega_{{\rm c}}}{J_{{\rm d}}} \;-\; 
\frac{cJ_{{\rm g}}a}{q|v_{\|}|}\; \pd{\omega_{{\rm c}}}{s} \;+\; \frac{c|p_{\|}|}{q}\;\pd{a}{s} \right). \label{eq:omegad_av} 
\end{equation} 
Using the identity 
\[ \frac{c|p_{\|}|}{q}\;\pd{a}{s} \;=\; \pd{}{s} \left( \frac{c}{q}\;|p_{\|}|a \right) \;-\; \frac{ca}{q}\;\pd{|p_{\|}|}{s} \;=\; \pd{}{s} 
\left( \frac{c}{q}\;|p_{\|}|a \right) \;+\; \frac{cJ_{{\rm g}}a}{q|v_{\|}|}\;\pd{\omega_{{\rm c}}}{s}, \] 
which follows from the definition (\ref{eq:p_par}) for $p_{\|}$, we find that the second and third terms in Eq.~(\ref{eq:omegad_av}) cancel each other and, hence, only the first term in Eq.~(\ref{eq:omegad_av}) remains. The bounce-averaged drift frequency is, therefore, given as 
\begin{equation} 
\langle \omega_{{\rm d}}\rangle \;=\; \frac{1}{\tau_{{\rm b}}} \sum_{\sigma}\;\int ds\; \left( \frac{J_{{\rm g}}}{|v_{\|}|}\;\pd{\omega_{{\rm c}}}{J_{{\rm d}}} \right) 
\;=\; J_{{\rm g}} \left\langle \pd{\omega_{{\rm c}}}{J_{{\rm d}}} \right\rangle. \label{eq:omegad_average} 
\end{equation} 
Based on this expression, we defined in Eq.~(\ref{eq:omegad_0}) the equatorial drift frequency $\omega_{{\rm d}0}$. 

An alternative expression for $\langle\omega_{{\rm d}}\rangle$ is obtained by introducing the bounce action 
\begin{equation} 
J_{{\rm b}} \;=\; \sum_{\sigma}\;\int \frac{ds}{2\pi}\;|p_{\|}| 
\label{eq:J_b} 
\end{equation} 
and the bounce frequency 
\begin{equation} 
\omega_{{\rm b}} \;=\; \frac{2\pi}{\tau_{{\rm b}}} \;=\; \left( \pd{J_{{\rm b}}}{\varepsilon} \right)^{-1}, \label{eq:omega_b} 
\end{equation} 
so that the bounce-averaged drift frequency (\ref{eq:omegad_average}) can also be expressed as 
\begin{equation} 
\langle\omega_{{\rm d}}\rangle \;=\; \;=\; -\;\frac{1}{\tau_{{\rm b}}} \sum_{\sigma}\;\int ds\; \pd{|p_{\|}|}{J_{{\rm d}}} \;=\; -\;\omega_{{\rm b}}\; 
\pd{J_{{\rm b}}}{J_{{\rm d}}}, 
\end{equation} 
where Eq.~(\ref{eq:p_par}) was used for $p_{\|}$. 

\vfill\eject


\begin{thebibliography}{99} 

\bibitem{Kaufman} A.N.~Kaufman, Phys.~Fluids {\bf 14}, 387 (1972). 

\bibitem{schulz74} M.~Schulz and L.J.~Lanzerotti, {\it Particle Diffusion in the Radiation Belts}, Springer-Verlag, New York, (1974). 

\bibitem{mynick89} 
H.~E. Mynick and R.~E. Duvall, 
% A unified theory of tokamak transport via the generalized {Balescu-Lenard} collision operator, 
Phys. Fluids {\bf B1}, 750 (1989). 

\bibitem{hudson98} 
M.~K. Hudson, V.~A. Marchenko, I.~Roth, M.~Temerin, J.~B. Blake, and M.~S. Gussenhoven, 
% Radiation belt formation during sudden storm commencements and loss during main phase, 
Adv. Space Res. {\bf 21}, 597 (1998). 

\bibitem{elkington99} 
S.~R. Elkington, M.~K. Hudson, and A.~A. Chan, 
% Acceleration of relativistic electrons via drift-resonant interaction with toroidal-mode {Pc-5 ULF} oscillations, 
Geophys. Res. Lett. {\bf 26}, 3273 (1999). 

\bibitem{elkington03} 
S.~R. Elkington, M.~K. Hudson, and A.~A. Chan, 
% Resonant acceleration and diffusion of outer zone electrons in an asymmetric geomagnetic field, 
J.~Geophys. Res. {\bf 108}, 1116, doi: 10.1029/2001JA009202 (2003). 

\bibitem{lyons71} 
L.~R. Lyons, R.~M. Thorne, and C.~F. Kennel, 
% Electron pitch angle diffusion driven by oblique whistler-mode turbulence, 
J.~Plasma Phys. {\bf 6}, 589 (1971). 

\bibitem{lyons74} 
L.~R. Lyons, 
% Pitch angle and energy diffusion coefficients from resonant interactions with ion-cyclotron and whistler waves, 
J.~Plasma Phys. {\bf 12}, 417 (1974). 

\bibitem{albert99} 
J.~M. Albert, 
% Analysis of quasi-liner diffusion coefficients, 
J.~Geophys. Res. {\bf 104}, 2429, (1999). 

\bibitem{horne03a} 
R.~B. Horne, S.~A. Glauert, and R.~M. Thorne, 
% Resonant diffusion of radiation belt electrons by whistler-mode chorus, 
Geophys. Res. Lett. {\bf 30}, doi: 10.1029/2003GL016963 (2003). 

\bibitem{BC_2001} A. J.~Brizard and A. A.~Chan, Phys.~Plasmas {\bf 8}, 4762 (2001). 

\bibitem{tsai84} 
S.~T. Tsai, J.~W. van Dam, and L.~Chen, 
% Linear relativistic gyrokinetic equation in general magnetically confined plasmas, 
Plasma Phys. Controlled Fusion {\bf 26\/}, 907 (1984). 

\bibitem{chen99} 
L. Chen, 
% Theory of plasma transport induced by low-frequency hydromagnetic waves, 
J.~Geophys. Res. {\bf 104\/}, 2421 (1999). 

\bibitem{BC_1999} A.J.~Brizard and A.A.~Chan, Phys.~Plasmas {\bf 6}, 4548 (1999). 

\end{thebibliography}
\end{document}